# Electrical System Architecture for Aviation Electrification

Anoy Saha and Mona Ghassemi

Department of Electrical and Computer Engineering, The University of Texas at Dallas

Richardson, TX 75080 USA

anoy.saha@utdallas.edu; mona.ghassemi@utdallas.edu

**Abstract:**
The electrification of aircraft is reshaping the foundations of aerospace design by positioning electrical systems at the center of propulsion, control, and onboard functionality. This chapter provides an overview of electrical system architectures for electric and hybrid electric aircraft, highlighting both established principles and emerging design strategies. The discussion begins with the motivations for electrification, including reducing environmental impact, improving operational efficiency, and replacing complex pneumatic and hydraulic subsystems with lighter and more reliable electrical alternatives. Aircraft electrical architectures are classified into four major categories: conventional, more electric, all electric, and hybrid electric. A range of system topologies is examined, including direct current (DC), alternating current (AC), hybrid, and distributed configurations. Each is considered in terms of its effectiveness in delivering power, enabling redundancy, supporting fault isolation, and managing thermal performance. Real world examples are presented to demonstrate practical applications, with case studies drawn from the Boeing 787 Dreamliner, the Eviation Alice commuter aircraft, and NASA's X57 Maxwell demonstrator. These examples illustrate the ongoing transition from incremental subsystem electrification toward fully integrated architectures that promise higher efficiency and greater sustainability.

**Keywords:** aircraft electrification; electric and hybrid aircraft; electrical system architectures; power distribution; all-electric propulsion; thermal performance.

## 1. Introduction

The transition to electric and hybrid-electric aircraft is reshaping the design and construction of onboard power systems, although hydrogen-fueled aircraft are considered a competitor [1]. However, the focus of this book chapter is on electrified aircraft. Electrical system architecture refers to the structured design and arrangement of components that handle the generation, conversion, distribution, and control of electrical energy within an aircraft. In current aircraft development, this architecture is no longer a background utility. It is now at the center of the platform's functionality, directly supporting propulsion, control, environmental systems, and safety operations [2], [3]. This transformation is driven by the aviation sector's need to reduce environmental impact, improve energy efficiency, and simplify system complexity [4], [5]. Traditional aircraft use a mix of mechanical, pneumatic, and hydraulic systems, all powered indirectly by fuel-driven engines. These systems are complex, heavy, and difficult to maintain. Electrical alternatives reduce overall system weight, allow for modular integration of components, and simplify maintenance schedules [4], [6], [7]. As a result, the electrical system architecture of an aircraft has become a primary focus in design.

In conventional aircraft, electrical systems were secondary. Most aircraft systems were powered using engine bleed air and hydraulic circuits, while electricity was mainly reserved for avionics and lighting. Power was delivered through generators running at 115 V and 400 Hz, connected in simple radial or split-bus configurations. This setup worked but had drawbacks. Bleed air extraction reduced engine efficiency and introduced thermal losses. Hydraulic lines required extensive mechanical routing, adding weight and increasing the risk of leaks or failures. Maintenance and operational costs were also high [5], [8].

In response, the concept of the more electric aircraft (MEA) was developed. Here, pneumatic and hydraulic systems are replaced with electrical equivalents. Environmental control, braking, and various actuation functions are now powered electrically. Aircraft such as the Boeing 787 and Airbus A350 adopt this approach, relying on high voltage AC and DC networks to supply a wider range of systems [2], [9]. Airbus has also explored hybrid-electric propulsion through its E-Fan X demonstrator program, further underscoring the growing industry shift [10]. This transition enhances overall efficiency and contributes to reduced emissions. The next stage is the all-electric aircraft (AEA), where electric systems power not only support equipment but also the propulsion system. In these aircraft, the electrical architecture must deliver power to high-performance motors, battery systems, and advanced control electronics. The design must accommodate rapid load changes, protect against faults, and manage thermal output, all while remaining as light and compact as possible [3], [11].

The design of electrical system architecture spans multiple domains. These include power sources (engines, batteries, fuel cells), conversion devices (inverters, converters, rectifiers), distribution systems (cables, busbars, connectors), protective elements (fuses, relays, sensors, and circuit breakers), and control units (power management and diagnostics systems) [3], [4]. Figure 1 shows the functional block diagram of aircraft electrical system architecture. Unlike ground-based systems, aircraft power systems must operate reliably under changing environmental conditions, including high altitudes, temperature extremes, vibration, and low pressure. The system must quickly detect and isolate faults to maintain stability. Redundant pathways are often included to provide backup in case of failure. Modern aircraft also operate at higher voltages to reduce current levels and decrease conductor size. Traditional systems use 28 V DC or 115 V AC. However, many new platforms are moving toward 270 V or higher DC systems. These reduce weight and improve efficiency but require special attention to insulation, thermal management, and protection from electrical faults [3], [6].

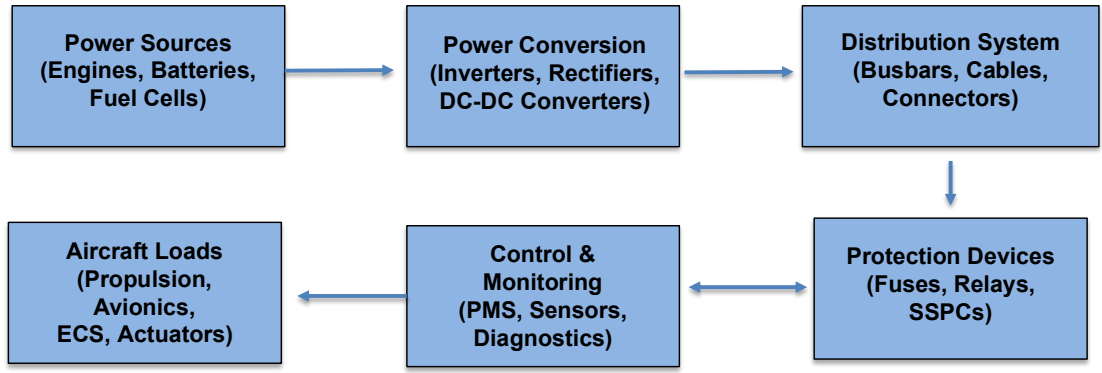

**Figure 1:** Functional block diagram of aircraft electrical system architecture.

This chapter presents a complete overview of electrical system architecture for electric and hybrid-electric aircraft. Section 2 outlines core design principles such as weight reduction, efficiency, and system reliability. Section 3 classifies electrical architecture across conventional, more electric, hybrid, and fully electric platforms. Section 4 explores system topologies including DC, AC, hybrid, and distributed approaches. Section 5 describes common power distribution networks and their configurations. Section 6 provides real-world examples from aircraft programs, and Section 7 concludes with a summary of key points.

## 2. Design Principles for Aircraft Electrical Systems

As aircraft increasingly rely on electrical systems for propulsion, control, and onboard operations, the way these systems are designed must evolve accordingly. In conventional aircraft, electrical power was mainly used for lighting, avionics, and communications, while key systems such as actuation and climate control were powered hydraulically or pneumatically. However, in modern electric and hybrid-electric aircraft, electricity plays a central role, powering everything from motors to environmental systems. This shift calls for a new set of design principles that ensure electrical systems are not only efficient and lightweight, but also safe, reliable, and ready for certification. This section explores the core ideas that guide the design of these systems, focusing on weight reduction, energy efficiency, system reliability, and regulatory compliance.

### 2.1 Weight Reduction

Weight is always a critical factor in aircraft design. In electric aircraft, the electrical system plays a major role in determining overall mass, especially through wiring, energy storage units, and power conversion equipment. Reducing this weight improves both range and efficiency and can also allow for increased payload or smaller energy storage capacity. One of the most effective ways to reduce weight in an electrical system is to increase system voltage. When voltage is increased, the same amount of power can be delivered with less current. A lower current makes it possible to use thinner cables, which are lighter and require less space. For example, switching from a 28 V system to a 270 V system can reduce the required wire thickness by a factor of ten or more, depending on the load [12], [13]. This reduction has a noticeable effect on total system weight, particularly in large aircraft where cable lengths are long. This approach reduces not only the mass of cabling but also structural support and routing complexity. In addition, combining several functions into a single compact unit, such as integrating sensing, control, and protection features into the power distribution hardware, helps reduce the total number of components and the wiring needed to connect them [8], [13].

**2.2 Energy Efficiency**

Energy efficiency is essential to extend range and maximize the performance of onboard power systems. Inefficiencies at any stage, whether during generation, conversion, or distribution, increase the energy required and may limit the effectiveness of electric propulsion. Designers address this by improving the efficiency of each component in the power chain. High-efficiency electric machines, optimized switching in power converters, and minimal loss in distribution wiring all contribute to better overall performance [14], [15]. Materials such as silicon carbide and gallium nitride are increasingly used in power electronics because they enable higher switching speeds, lower conduction losses, and reduced thermal output [16]. Efficiency is also improved by reducing conversion steps and simplifying the path from energy source to load. Shorter routing, DC distribution, and intelligent power management reduce unnecessary losses and help stabilize system behavior under dynamic conditions [16].

**2.3 Reliability and Redundancy**

Aircraft electrical systems must operate reliably under a wide range of flight conditions and respond safely to faults. Any loss of power in critical systems must be avoided, and architecture must include measures to detect, isolate, and recover from failures. Reliability begins at the component level with robust design, thorough testing, and environmental qualification [8], [17]. At the system level, redundancy ensures that power is always available to essential loads. This includes multiple independent power sources, separate distribution paths, and parallel converter units. Configurations such as dual-bus and multi-bus networks are used to prevent a single fault from affecting the entire system. Real-time monitoring is also important. Sensors continuously track voltage, current, and temperature at key points in the system. If anomalies are detected, protection devices can disconnect faulty sections and reroute power as needed [17], [18]. In electric propulsion systems, this capability is essential for maintaining safe operation throughout the flight envelope.

**2.4 Regulatory and Safety Standards**

All aircraft electrical systems must comply with international safety and performance regulations. These standards ensure that systems are designed and operated within safe limits, especially under fault or abnormal conditions. Agencies such as the Federal Aviation Administration (FAA) and the European Union Aviation Safety Agency (EASA) provide comprehensive guidelines for certification and airworthiness. Relevant standards include DO-160 for environmental testing [19], DO-254 for airborne electronic hardware design assurance [20], and ARP4754A for system development processes [21]. These documents specify procedures for component testing, electromagnetic compatibility, fault isolation, and system-level verification. They are especially important when developing systems that include high-voltage equipment, lithium-based energy storage, and electric propulsion. In high-voltage systems, additional safety requirements apply. These include isolation monitoring, arc fault protection, thermal runaway mitigation, and appropriate clearance and creepage distances [22]. The growing use of DC at high voltages introduces new challenges that must be addressed in both design and certification.

## 3. Classification of Aircraft Electrical Architectures

As aviation technology evolves, the way electrical systems are used in aircraft is changing rapidly. Based on how much the aircraft depends on electricity, electrical system architectures can be grouped into four main types: conventional, more-electric, all-electric, and hybrid-electric. These categories help us understand how aircraft have developed from using mechanical systems to becoming increasingly powered by electricity.

### 3.1 Conventional Aircraft Electrical Systems

In traditional aircraft, electricity is used in a very limited way. As shown in Figure 2, the main source of energy for most onboard systems is mechanical, hydraulic, or pneumatic. Hydraulic systems use pressurized fluid to move components such as landing gear and flight control surfaces. Pneumatic systems use air pressure, usually taken from the engine, to operate systems like de-icing and air conditioning. In contrast, the electrical system in these aircraft only powers basic items like cockpit instruments, lighting, entertainment systems, and communication devices.

The electrical power in conventional aircraft is typically provided by generators connected to the engines. These generators produce constant-frequency AC at 115 V and 400 Hz, along with 28 V DC for low-voltage systems [3]. To keep the frequency constant even when the engine speed changes, a special device called a constant-speed drive is used. These systems are reliable but heavy and complex, and they require regular maintenance.

While this setup has worked well for decades, it is not very efficient. The many mechanical parts, long pipe and cable networks, and engine bleed systems add weight and reduce overall performance. These limitations led to the development of the next generation of electrical systems.

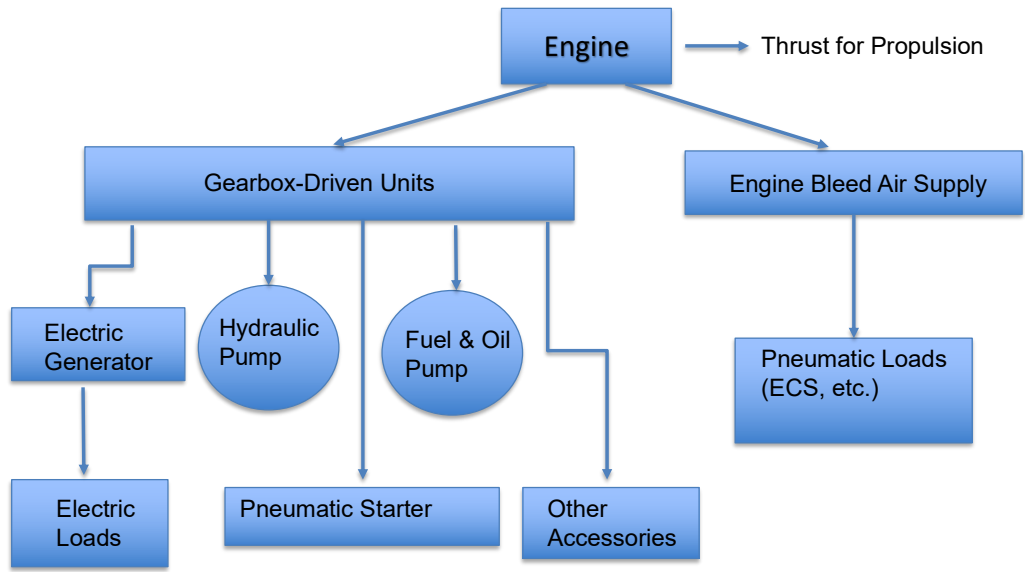

**Figure 2:** Conventional aircraft electrical system architecture.

**3.2 More-Electric Aircraft (MEA) Systems**

More electric aircraft represent the first major step toward replacing mechanical systems with electrical ones. In MEA designs, many of the components that were previously powered by hydraulic or pneumatic systems are now electrically driven. For example, instead of using hydraulic fluid to move wing flaps or control surfaces, MEAs use electric motors called electro-mechanical actuators. Electric fuel pumps, electric brakes, and even electrically driven compressors for cabin air are common in MEAs [23].

This shift improves overall efficiency, reduces weight and maintenance costs. It also allows for easier system integration and automation. However, since these systems require more power, MEAs need more advanced electrical architecture. In conventional aircraft, power requirement might be around 250 to 400 kVA, but in MEAs it can exceed 1 MVA [5].

To support this demand, MEAs often use both AC and DC systems at higher voltages. As shown in Figure 3, some aircraft use variable-frequency AC (360–800 Hz) and ±270 V DC systems. There are several common architecture types referred to as EPS-A1 through EPS-A4 [5], [8]. The Boeing 787 is a leading example of an MEA. It uses electric power for many functions that were previously hydraulic or pneumatic, such as de-icing, brakes, and cabin pressurization [23].

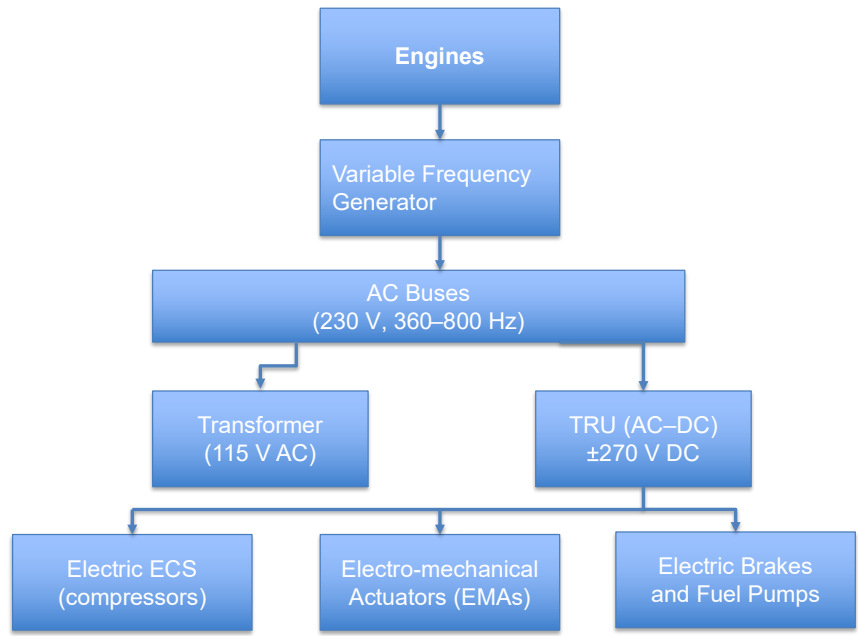

**Figure 3:** Schematic of MEA Power Distribution.

### 3.3 Hybrid-Electric Aircraft Systems

Hybrid electric aircraft systems combine conventional fuel-based engines with electric components, using the engines both for propulsion and to charge onboard batteries that can supplement power during flight [24]. They operate in three configurations, see Figure 4:

- **Series Hybrid**: In a series hybrid system, the fans are powered solely by electric motors, while the traditional engine operates a generator that supplies electricity to the motors and can also charge the onboard batteries.
- **Parallel Hybrid**: In a parallel hybrid system, both a battery-powered motor and a conventional fuel engine are connected to the same shaft that drives the fan, allowing either or both to deliver propulsion depending on the flight condition.
- **Series-Parallel Hybrid**: In a series/parallel partial hybrid system, some fans are mechanically powered by a conventional fuel-based engine, while others are driven only by electric motors supplied with power from either a battery or a generator connected to a turbine.

Hybrid architectures offer fuel savings, noise reduction during takeoff/landing, and backup power. As battery energy density improves, hybrid-electric aircraft may become more prevalent for medium-range flights where fully electric systems remain impractical.

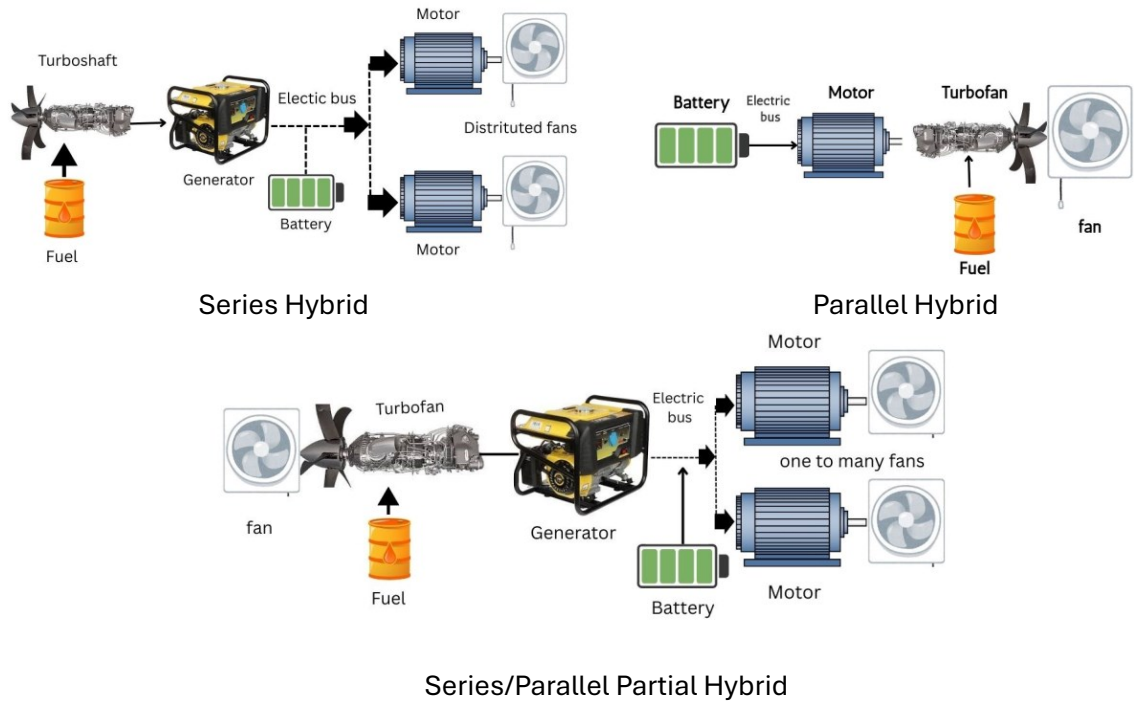

Series Hybrid  Parallel Hybrid

Series/Parallel Partial Hybrid

**Figure 4:** Schematic of Hybrid Electric Aircraft Systems.

**3.4 All-Electric Aircraft (AEA) Systems**

AEA represents the most advanced stage of aircraft electrification, where all onboard systems, including propulsion, are powered exclusively by electricity. These aircraft eliminate gas turbines, bleed air systems, and fuel-based engines entirely. Instead, they rely on electrochemical energy units (EEUs) such as batteries, fuel cells, or supercapacitors as the sole source of energy [25, 26]; see Figure 5.

The complete removal of hydraulic and pneumatic subsystems results in a simplified aircraft architecture, improved maintainability, and zero in-flight emissions. However, this high level of electrification poses significant challenges. Propulsive loads for large aircraft can exceed 25 MW, necessitating the use of compact, lightweight, and high-power energy storage systems. To manage efficiency and conductor sizing, AEA employs high-voltage systems, often above 500 V, with medium-voltage direct current (MVDC) systems reaching ±5 kV.

A prime example is the envisaged NASA N3-X all-electric aircraft, which incorporates four EEUs powering 14 distributed electric motors, each rated around 1.785 MW. The proposed electrical power system supports ±5 kV MVDC buses for propulsion and secondary ±0.5 kV and 28 V DC buses for non-propulsive loads such as avionics and environmental control systems (ECS). Different EPS architectures have been proposed and evaluated under failure conditions, including EEU or cable loss, to validate system-level fault tolerance [25]. DC architectures are favored over AC due to better efficiency, lower weight, and direct compatibility with battery and fuel cell outputs. Comparative studies indicate that ±5 kV DC systems outperform 10 kV AC alternatives in mass and loss reduction for large-scale aircraft applications.

While the envisaged all-electric N3-X in [25] highlights future capabilities, current AEA is more practical for short-haul or low-capacity missions. The Pipistrel Velis Electro, certified by EASA, is designed for flight training and short-range travel using lithium-ion batteries [26]. The Eviation Alice targets regional commuter routes up to 300 miles with nine-passenger capacity and fully battery-powered propulsion [27].

Urban air mobility platforms are also emerging. The Lilium Jet and Airbus CityAirbus NextGen are electric vertical takeoff and landing (eVTOL) aircraft designed for intra-city transport. These aircraft use multiple small electric fans or ducted propellers to achieve vertical lift and efficient cruise, powered entirely by onboard batteries [28], [29].

As energy storage technologies evolve, future AEA may scale up in range and payload. Continued advancements in battery energy density, fuel cell power-to-weight ratios, and MVDC system integration are key to unlocking the full potential of emission-free aviation [5].

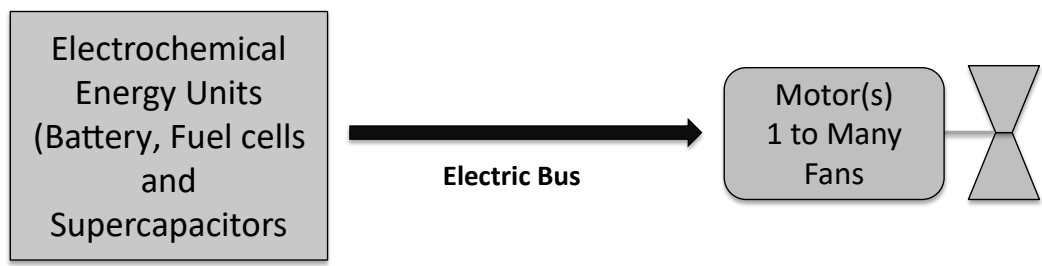

**Figure 5:** Schematic of All-Electric Aircraft Systems.

## 4. Electrical System Topologies

Modern electric and hybrid-electric aircraft employ several distinct electrical distribution topologies. The choice of topology depends on performance goals such as minimizing weight, ensuring safety, improving efficiency, and simplifying maintenance. The four main categories are DC distribution, AC distribution, hybrid AC/DC systems, and distributed power systems. Each topology reflects trade-offs in efficiency, fault tolerance, and compatibility with loads and sources.

**4.1 DC Power Distribution**

DC power distribution has become increasingly attractive for more electric and AEA because it offers higher efficiency and reduced weight compared to AC systems. A typical configuration, as shown in Figure 6, uses ±270 V DC buses, providing 540 V between the positive and negative lines. This higher voltage reduces current for a given power level, which in turn allows the use of lighter and thinner conductors, lowering both resistive losses and overall wiring weight [25], [30].

Aircraft such as the Boeing 787 introduced ±270 V DC buses for systems with high electrical demand, including wing ice protection system (WIPS), cabin compressors, and actuators [31]. Airbus has also explored these systems in research programs such as Clean Sky, which developed a 270 V DC distribution unit [32]. Since batteries and fuel cells produce DC naturally, these systems can connect directly without conversion. However, DC systems must deal with issues like arc suppression and insulation challenges, especially in low-pressure flight environments [33], [34].

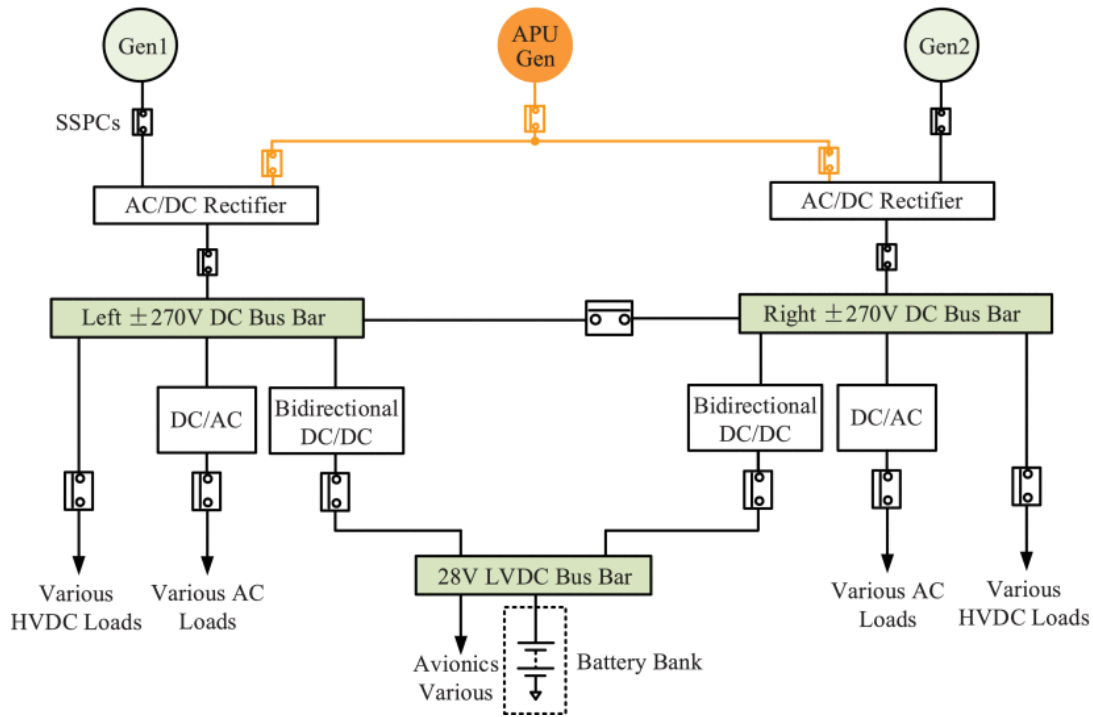

**Figure 6:** Schematic of DC Power Distribution [30].

**4.2 AC Power Distribution**

AC has been the traditional method for distributing electrical power in aircraft, see Figure 7. Early designs commonly used 115 V, 400 Hz systems because the high frequency allowed smaller and lighter transformers. These constant-frequency systems used constant-speed drives (CSDs) to maintain 400 Hz output despite engine speed variations. Modern designs, such as the Airbus A380 and Boeing 787, transitioned to variable-frequency AC (360–800 Hz), eliminating the heavy CSD while introducing electronic loads capable of handling frequency variation.

The Boeing 787 also raised AC voltage to 230 V variable-frequency to reduce current and wiring losses. AC power remains valuable for lighting, avionics, and cabin systems. However, AC networks demand generator synchronization, can suffer from reactive power losses, and require more complex protection. They are also heavier, since three conductors are required for three-phase systems, compared with two for DC [30].

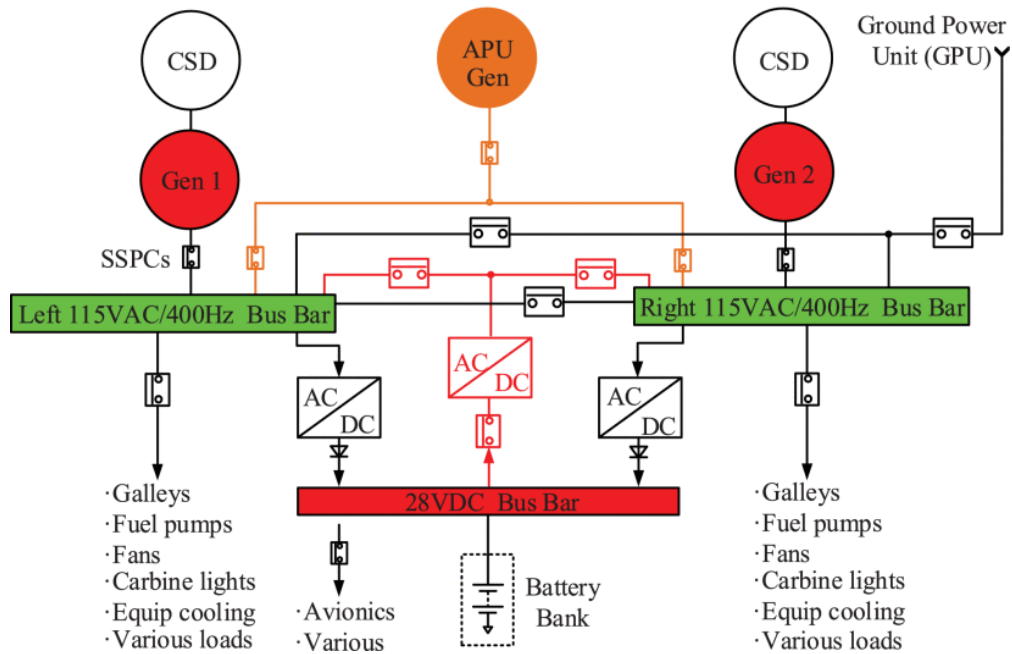

**Figure 7:** Schematic of AC Power Distribution [30].

**4.3 Hybrid AC and DC Systems**

Many modern aircraft employ a hybrid electrical architecture that combines both AC and DC distribution to meet the diverse requirements of onboard systems. In this configuration shown in Figure 8, two main generators along with an auxiliary power unit (APU) generator supply variable-frequency AC power (115 VAC, 360–800 Hz). This power is distributed through left and right AC bus bars, which then feed different conversion units. Some AC is used directly by variable-frequency loads, while AC/AC converters provide constant-frequency (115 VAC, 400 Hz) power for systems that require stable operation. Transformer rectifier units (TRUs) and auto-transformer rectifier units (ATRUs) convert AC into DC, creating 270 VDC buses for high-voltage loads and 28 VDC buses for avionics and low-voltage systems, with battery banks providing backup. This combination of AC and DC networks allows flexible power sharing, supports both high-voltage and low-voltage demands, and ensures redundancy for critical aircraft functions.

The Airbus A380 uses both 115 V AC and 270 V DC buses to power systems with different requirements [30]. The Boeing 787 takes this concept further with 230 V variable-frequency AC and ±270 V DC. This allows both traditional and high-power electrical systems to operate together in a coordinated way. These setups are useful for load balancing. For example, batteries can be connected to the DC bus to help during takeoff or to provide emergency power. Inverters and converters allow power to move between AC and DC buses when needed [33].

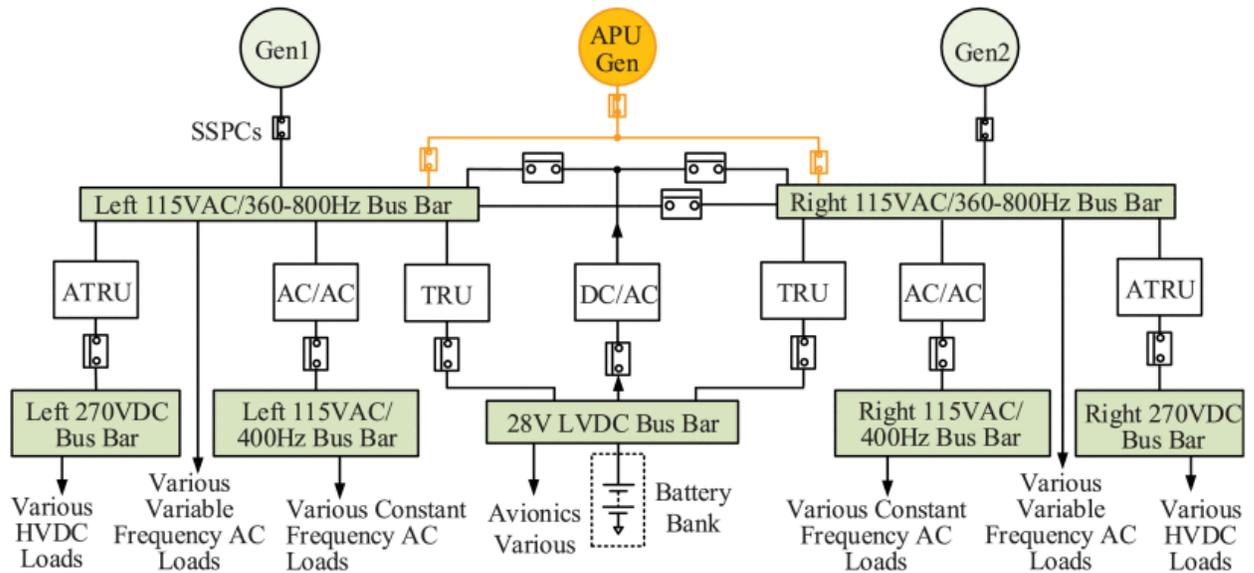

**Figure 8:** Schematic of Hybrid AC/DC systems [30].

**4.4 Distributed Power Systems**

Distributed power systems move power electronics closer to the loads, reducing cabling weight and improving fault isolation. This is particularly important for distributed electric propulsion (DEP) aircraft, where multiple motors are placed across the wing or fuselage. Each motor has its own inverter and controller located nearby. This reduces the length of heavy cables and allows better fault isolation. If one motor fails, others can still work. NASA's X-57 Maxwell uses this design with 12 high-lift motors along the leading edge of the distributed electric propulsion wing. Each motor has its own electronics to control speed and torque [36]. Other aircraft, such as the ECO-150 and Boeing SUGAR Volt, also use distributed power systems that allow more flexibility and redundancy [37], [38].

## 5. Electrical Power Distribution Networks

The power system architectures discussed in Sections 4.1–4.3 are primarily intended for MEA, where the required power is relatively small, around 1 MVA, compared to wide-body or large AEA, which demand over 25 MW for takeoff. Distributing such a large amount of power across multiple motors/fans and generators necessitates medium-voltage distribution in the few-kV range, which is the focus of this section.

In an all-electric aircraft, the generation of power is only the first step in enabling the operation of propulsion motors, actuation systems, environmental control equipment, and the many auxiliary functions needed in flight. Once the energy is produced, it must be distributed throughout the aircraft in a way that is both efficient and highly reliable. The distribution network is therefore one of the most critical elements of the overall electrical system. Its configuration determines how the aircraft responds to faults, how quickly power can be restored after an interruption, how much wiring is required, and how maintenance is carried out. Engineers must carefully balance weight, safety, and operational flexibility when selecting appropriate architecture. The main distribution approaches used in aircraft are the ring bus, the radial system, the dual or multi-bus system, and the direct current microgrid.

### 5.1 Radial Bus Configurations

A radial bus is the simplest way to distribute power in an aircraft. Power flows outward from the source to each load along a single path, much like branches extending from a tree. This simplicity results in fewer components, lower system mass, and easier design. However, it also creates a vulnerability: if one feeder is damaged or develops a fault, all loads downstream lose power immediately. To improve resilience, designers often divide the system into multiple radial branches and add tie switches that can reconnect isolated sections during an emergency. Even with these improvements, the radial design provides less redundancy than more complex configurations, which limits its application to smaller aircraft or non-critical systems where simplicity and weight savings are more important than fault tolerance.

NASA's X-57 Maxwell demonstrator initially employed a radial HVDC distribution for its distributed electric propulsion system: essentially four separate DC buses, each supplying a set of motors on the wing [39]. This arrangement kept wiring short and straightforward, reducing weight. However, the drawback was clear: if one bus failed, all motors connected to it would shut down. On a wing with many motors, such a failure could cause severe asymmetric thrust and compromise flight stability. Later design phases of the X-57 introduced greater redundancy to mitigate this risk, but the case highlights the trade-off inherent in radial systems. They are efficient and lightweight but require careful planning to avoid unacceptable consequences of a single fault.

### 5.2 Dual-Bus and Multi-Bus Configurations

Dual-bus and multi-bus systems are designed to balance redundancy and weight. In a dual-bus arrangement, the aircraft has two main power channels, each fed by its own generator or battery. Under normal conditions the buses operate independently, supplying different groups of loads. If one generator or bus fails, tie connections allow the healthy side to power both sets of loads, ensuring that no essential function is lost. Multi-bus systems expand on this principle by adding more buses, allowing finer control of which loads are powered during failures. This makes it possible to shed non-critical services and reserve power for propulsion, avionics, and flight controls. Compared with a ring bus configuration explained in Section 5.3, multi-bus systems are lighter and less complex, but they still provide strong resilience against failures, making them standard for large commercial aircraft.

The Boeing 787 Dreamliner and Airbus A380 provide real-world examples of multi-bus designs. The Boeing 787 uses four main AC buses, each connected to variable-frequency generators, and also produces ±270 V DC for subsystems. If one generator or bus fails, automated tie switches reroute power so that essential loads remain supplied. This architecture allows the aircraft to withstand multiple failures without losing critical systems. Similarly, the Airbus A380 employs four large engine-driven generators that supply separate AC buses, which can be cross-tied to share loads when needed. Experimental AEA also apply this principle.

**5.3 Ring Bus Configurations**

A ring bus architecture forms a closed-loop power network in which sources and loads are connected in a ring (loop). If a fault occurs on one part of the loop, the system can close that open segment and redirect power flow in the opposite direction, maintaining supply to the loads. This feature makes the ring bus one of the most fault-tolerant network designs, since each load can be supplied from two directions. However, the advantages come with trade-offs: more switches, longer cabling, and more complex protection and control schemes are required. These add weight and cost, which are critical concerns for aircraft. Despite these challenges, the ring bus remains attractive for large or safety-critical all-electric aircraft, where ensuring continuous propulsion power outweighs the penalties.

The NASA N3-X concept aircraft is a strong example of a ring bus in practice. The N3-X is a wide-body, hybrid wing-body design that uses multiple distributed electric fans for propulsion. In its proposed electrical power system, superconducting generators at the wingtips supply a medium-voltage DC ring. If one generator or a bus segment fails, contactors in the ring can reconfigure the network so that power is routed through the remaining healthy paths, keeping all 14 propulsion motors running. During a one-engine-out condition, the normally open ties in the loop close, allowing the remaining engine's generators to feed the entire fan array. This design demonstrates how a ring bus ensures that no single fault can disable propulsion, making it a promising option for future AEA where uninterrupted thrust is essential.

**5.4 MVDC Microgrid Approach**

A MVDC microgrid is the most advanced and flexible distribution architecture considered for all-electric aircraft. In this system, all sources and loads connect to a common high-voltage DC backbone. Power electronic converters manage energy flow in both directions, so batteries can charge during cruise and discharge during takeoff, while fuel cells or turbogenerators can supplement propulsion as needed. This integration allows the aircraft to function as a coordinated microgrid, balancing supply and demand dynamically. The DC approach reduces conversion steps, saves weight compared to AC systems, and works naturally with modern storage technologies. The main challenge is protection. Because DC currents do not cross zero, faults are harder to interrupt, requiring specialized solid-state breakers, arc-fault detection, and very fast control. Despite these hurdles, the MVDC microgrid is increasingly viewed as the most promising long-term solution for future large electric aircraft.

NASA's N3-X all-electric aircraft concept [25, 26] showcases the feasibility of employing a medium-voltage DC (MVDC) microgrid. Its architecture utilizes ±5 kV DC buses to deliver a combined takeoff power of 25 MW across 14 distributed propulsion motors [25]. The backbone is supported by four electrochemical energy modules—a mix of batteries, fuel cells, and supercapacitors—coordinated to ensure both propulsion and auxiliary power requirements are continuously satisfied. Similar design philosophies have been explored in demonstrators such as Airbus's E-Fan X and NASA's STARC-ABL hybrid, which employed DC buses to couple turbogenerators with storage systems. These examples underscore how microgrid-based solutions enhance modularity, scalability, and efficiency in distributed propulsion, albeit while introducing significant challenges in protection and control.

Figure 9 presents a schematic of the N3-X electric power system (EPS) [25], highlighting its fully electric configuration. In the diagram, "M" denotes motors, with "R" and "L" indicating those on the right and left wings, respectively. To maintain clarity, the figure omits auxiliary components such as the APU and sensors, instead focusing on the propulsion subsystem—the dominant power consumer—and on non-propulsive demands, including the environmental control system (ECS), wing ice protection system (WIPS), electric taxiing, actuation, hotel and galley loads, lighting, and avionics. In [25], three MVDC architectures rated at ±5 kV (10 kVdc) were proposed for the propulsion system and benchmarked against equivalent 10 kVac AC EPSs. These three MVDC topologies are analyzed in detail below. For non-propulsive loads, a bipolar ±0.5 kVdc bus has been suggested, while avionics and other low-voltage systems can continue to operate on 28 Vdc supplies, consistent with conventional aircraft and currently commercialized MEA platforms such as the Boeing 787.

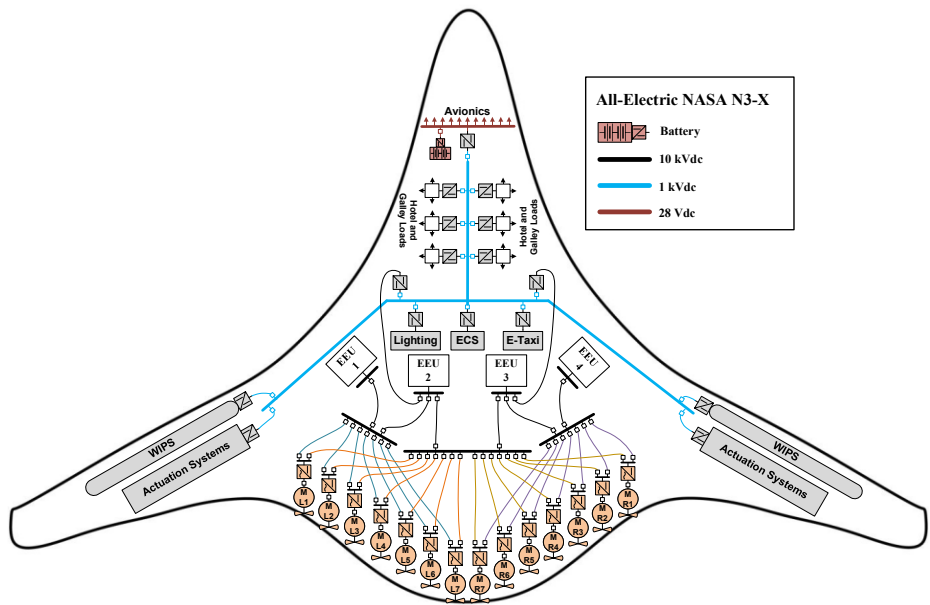

**Figure 9:** Electric power system proposed for N3-X NASA aircraft, imagined as all-electric [25].

For the NASA N3-X concept, the required thrust power at takeoff is estimated at 25 MW. To meet this demand, a ±5 kVdc MVDC electric power system (EPS) is proposed, supplying fourteen distributed propulsion motors, each rated at approximately 1.785 MW of constant power during takeoff [25]. Three all-electric propulsion architectures for this EPS were introduced in [25], illustrated in Figures 10, 11, and 12 as Architectures #1, #2, and #3, respectively.

In all three designs, four electrochemical energy units (EEUs)—comprising supercapacitors, batteries, and fuel cells—collectively provide thrust power to feed the fourteen motors. Each architecture is designed to ensure uninterrupted power delivery both under normal operating conditions and in single-contingency scenarios, including severe events such as the complete loss of one EEU. Although Figures 10–12 present DC-based configurations, these EPSs can also be interpreted as AC systems with equivalent single-line diagrams for the propulsion sections.

The analyses in [25, 26] confirm that each proposed EPS satisfies both power flow and bus voltage requirements under normal operation and across all n–1 contingencies. Contingencies considered include failures of busbars, cables (branches), and EEUs. The most critical events often involve the loss of a busbar or an EEU. For example, in Architecture #1, failure of bus #5 results in the loss of EEU1, the associated motors, and cables 1–5 and 2–5. Despite this, the event is treated as a single contingency—loss of one busbar. The EPS remains capable of sustaining power delivery to the motors by rerouting through alternative paths. Similarly, in the case of EEU failures, the most severe condition is the complete loss of a single EEU. If, for instance, EEU2 fails, it can no longer supply power; however, in Architecture #1, bus #2 remains active, allowing power to be transferred from bus #5 through bus #2 and then onward to bus #6, thereby maintaining motor operation.

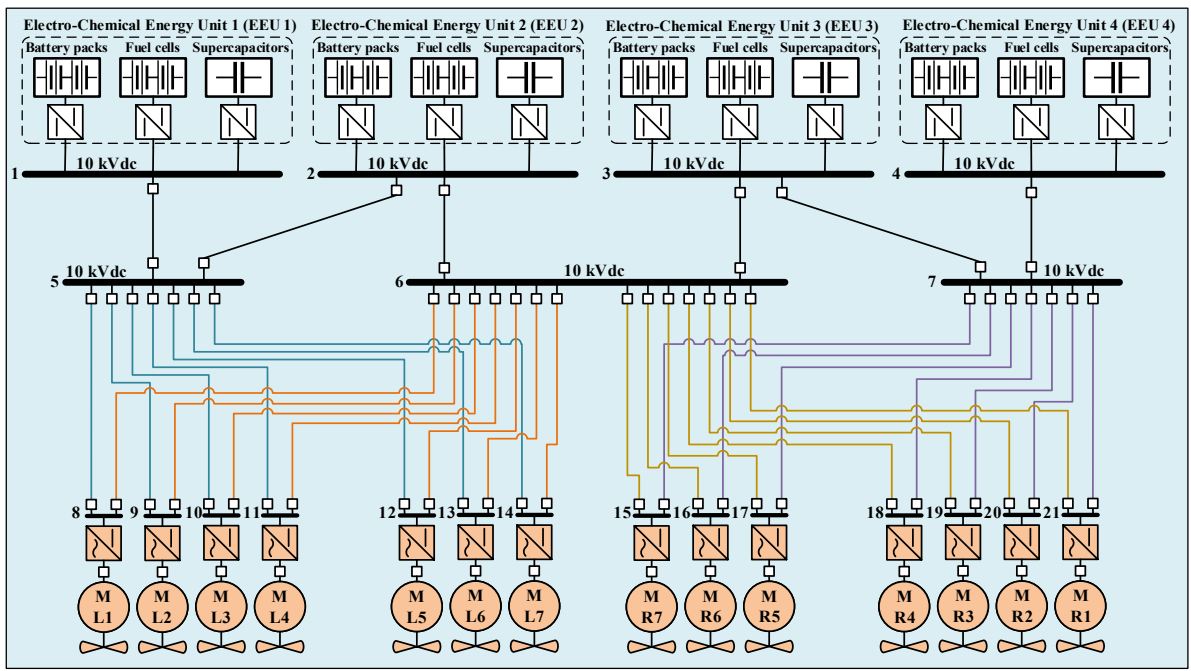

**Figure 10:** Propulsion section of the electric power system for NASA's N3-X all-electric aircraft, Architecture #1 [25]. MR1: motor #1 on the right wing; ML1: motor #1 on the left wing.

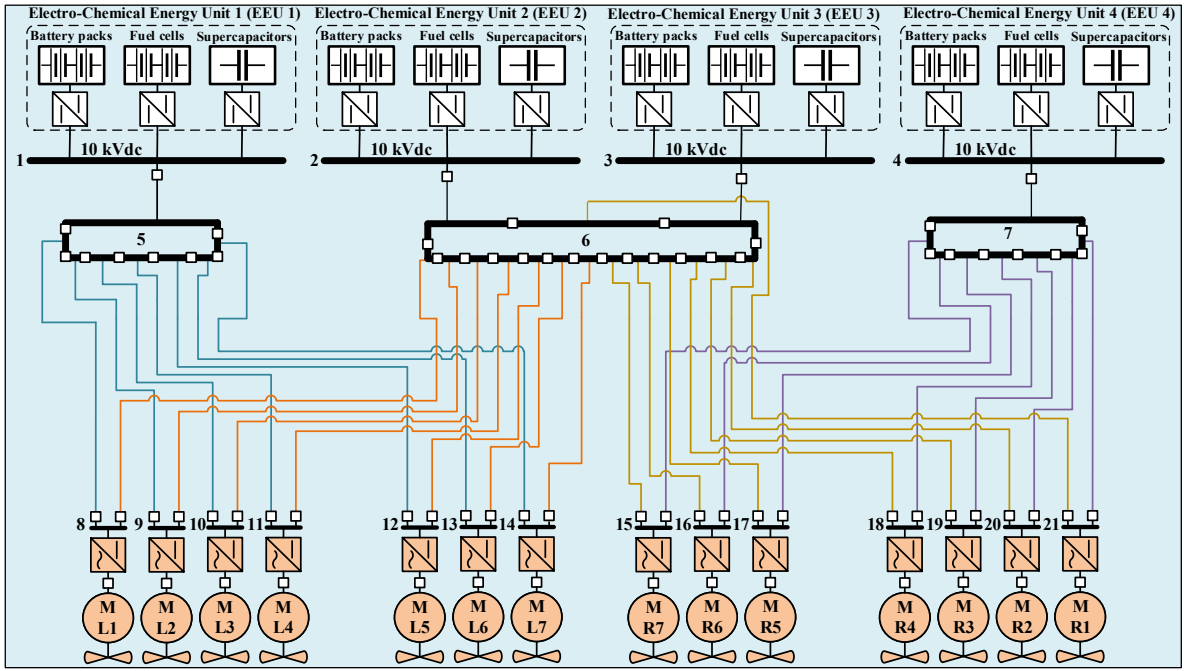

**Figure 11:** Propulsion section of the electric power system for NASA's N3-X all-electric aircraft, Architecture #2 [25].

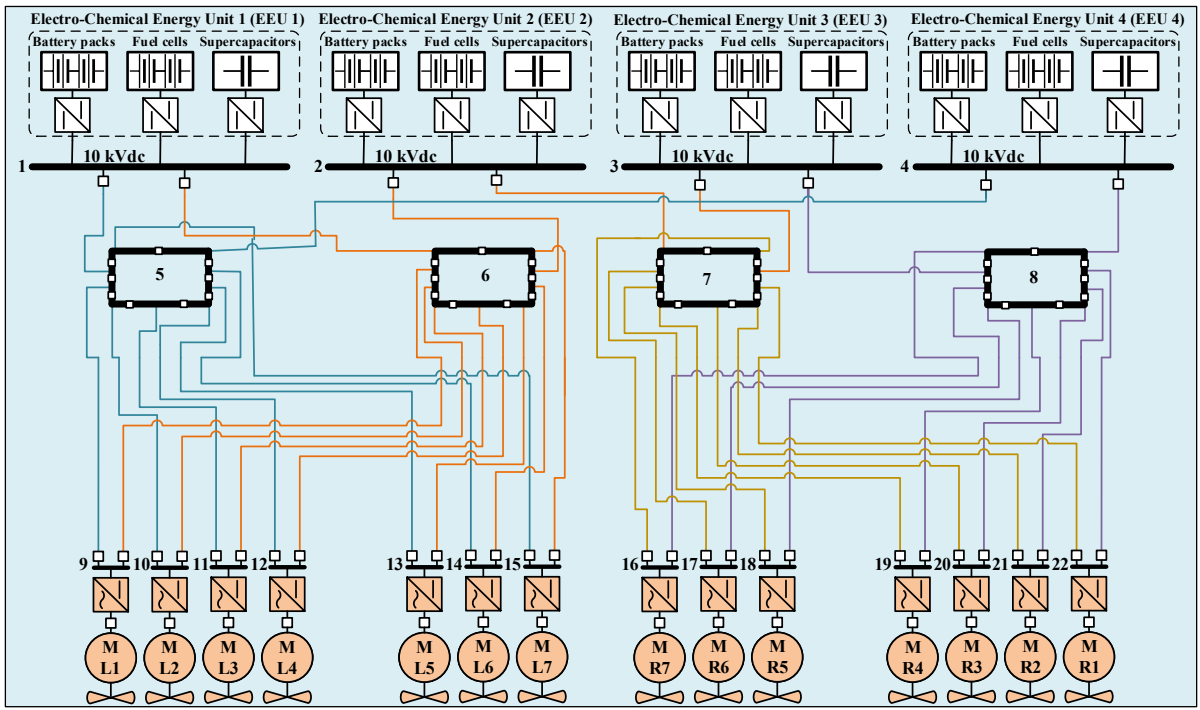

**Figure 12:** Propulsion section of the electric power system for NASA's N3-X all-electric aircraft, Architecture #3 [25].

The architectures shown in Figures 10–12 are the most sophisticated proposed to date. It is important to note that the design of MVDC (±5 kV), high-power (1 kA) cables required for these architectures faces significant thermal challenges due to limited heat transfer at cruising altitude (12.2 km), where the low-pressure environment exacerbates cooling issues [40], [41]. To address these challenges, several cables have been developed incorporating the novel concept of Micro Multilayer Electrical Insulation (MMEI) systems [42]-[52]. These cables are specifically engineered to suppress partial discharges (PDs) and arcs while supporting higher current densities. Experimental results further demonstrated that they maintain high breakdown strength even under low-pressure conditions, making them suitable for high-voltage applications in electric aircraft [53]-[61].

## 6. Case Studies of Existing Architectures
### 6.1 Boeing 787 Dreamliner (MEA)
The Boeing 787 Dreamliner is a prime example of the More-Electric Aircraft paradigm in commercial aviation. It eliminates traditional bleed-air pneumatic drives in favor of electrically powered systems for functions like cabin pressurization, wing ice protection, engine start, and even braking [62]. The 787's engines and APU together drive four generator units mounted on the two engines that supply over 1 MVA of electrical power, distributed via a mixed AC/DC architecture. Notably, engine-mounted Variable Frequency Generators deliver 235 VAC, which is rectified into a high-voltage ±270 V DC bus to run the largest loads such as the electrical environmental control compressors and electro-thermal wing de-icing heaters. The 787 also introduced all-electric brake actuators in its landing gear, replacing hydraulic lines with quicker and lighter electromechanical systems. Overall, the 787's integrated modular avionics and power management network demonstrate how replacing bleed air and hydraulics with electrics can yield efficiency gains (about 14% lower operating costs than its predecessor) while maintaining high reliability through power quality monitoring and solid-state protection devices [63].

### 6.2 Eviation Alice (All-Electric)
The Eviation Alice aircraft showcases a pure electric propulsion system, supported entirely by a centralized lithium-ion battery system. The battery pack, rated close to 900 kWh, operates at approximately 820 VDC and supplies two MagniX electric motors. The powertrain includes high-power DC-DC converters and motor inverters, each equipped with thermal management systems and digital controllers. Charging is managed via CCS-type fast chargers with onboard battery management systems ensuring thermal protection and balanced charging. The aircraft integrates lightweight power cables, high-voltage isolation relays, and thermal runaway detection systems. Its simplicity and modularity make Alice a pioneer in short-range electric commuter aviation [64], [65].

### 6.3 NASA X-57 Maxwell (Hybrid)
NASA's X-57 Maxwell is an experimental aircraft created to demonstrate DEP and new approaches to electric aircraft systems. It is a modified Tecnam P2006T that, in its final configuration (Modification IV), features 14 electric motors: 12 small high-lift motors with foldable propellers along the wing leading edge, plus 2 larger cruise motors at the wingtips. All motors are powered by two high-voltage lithium-ion battery packs, which together provide a nominal bus voltage of about 460 V DC (operating range ~320–540 V) [66].

**6.4 Airbus/Rolls-Royce E-Fan X (Hybrid-Electric Demonstrator)**

The E-Fan X was a high-profile hybrid-electric aircraft project by Airbus, Rolls-Royce, and Siemens, intended to test megawatt-scale electric propulsion in a regional jet. The demonstrator was based on a BAe 146 airliner, with one of its four turbofan engines removed and replaced by a 2 MW electric fan motor on the wing [67]. In this serial hybrid design, a Rolls-Royce AE2100 gas turbine engine (mounted in the rear fuselage) drove a 2.5 MW generator, which, together with a 2-ton lithium-ion battery pack in the cargo bay, would supply power to the electric motor.

## 7. Conclusion

The shift toward aircraft electrification is changing the way power systems are designed and integrated, moving them from supporting roles to the very core of propulsion and control. This chapter has shown how electrical architecture has progressed from conventional arrangements with limited electrical functions to more-electric aircraft that replace pneumatic and hydraulic systems with electrical alternatives, and finally toward hybrid-electric and all-electric platforms that place propulsion itself within the electrical domain. Each stage of this evolution has demanded new approaches to generation, conversion, distribution, and protection, and each has brought unique engineering challenges that must be addressed to ensure safety and reliability.

The comparison of different network topologies highlights the trade-offs that shape aircraft design. Radial systems remain attractive for their simplicity and light weight, but their limited fault tolerance restricts their use to smaller platforms or less critical loads. Dual and multi-bus systems provide a balance between redundancy and efficiency and are already the standard in modern commercial aircraft. Ring bus networks go further by offering alternate pathways for power flow, greatly improving reliability, though at the cost of added weight and complexity. MVDC microgrids represent the most forward-looking solution, integrating diverse energy sources such as batteries, fuel cells, and turbogenerators into a common backbone that supports distributed propulsion. Yet, these microgrids require new methods for fault detection, arc suppression, and high-voltage protection before they can be fully adopted in large-scale aviation.

The case studies illustrate how these architectures are being realized in practice. The Boeing 787 shows how the more-electric approach can reduce reliance on pneumatic systems and improve efficiency through high-voltage AC and DC networks. The Eviation Alice demonstrates the potential of a fully battery-electric commuter aircraft, while NASA's X-57 Maxwell highlights the promise of distributed propulsion and modular electrical subsystems. Hybrid demonstrators such as the Airbus E-Fan X prove the feasibility of megawatt-scale electrical power integration, and the NASA N3-X concept points toward a long-term future where superconducting machines and medium-voltage DC distribution enable wide-body all-electric flight. Together, these examples illustrate a clear trend of progressive electrification, moving step by step toward fully electric aircraft.

Although progress has been substantial, many challenges remain unresolved. Energy storage technologies must advance to provide higher specific energy and faster charging capabilities. Thermal management of motors, inverters, and cables must be improved to cope with the high power densities expected in flight. Protection strategies for high-voltage DC must mature so that faults can be detected and cleared safely at altitude. At the same time, new materials such as wide-bandgap semiconductors and superconductors offer pathways to greater efficiency and reduced weight, while intelligent power management systems are emerging to orchestrate complex electrical networks in real time.

Looking ahead, the development of electrical system architecture will remain central to the future of aviation. More-electric and hybrid-electric platforms provide essential transitional steps, while AEA represents the ultimate goal once enabling technologies have matured. As the aviation industry pursues ambitious environmental targets, the design of safe, reliable, and efficient electrical architecture will be a decisive factor in achieving sustainable flight. Electrical power is no longer an auxiliary service in aircraft; it is becoming the defining element of twenty-first century aerospace design.